# Topologically compensated polar skyrmions in twisted oxide membranes

Young-Hoon Kim[1*], Harrison A. Walker[2,3*], Haozhi Sha[4], Reza Ghanbari[5], Haoyang Ni[1], Eli Rodrigues[5], Jianwei Miao[4], Ramamoorthy Ramesh[6], Ruijuan Xu[5], Sokrates T. Pantelides[2,3,7*], Miaofang Chi[1,8*]

[1]Center for Nanophase Materials Sciences (CNMS), Oak Ridge National Laboratory (ORNL), Oak Ridge, TN 37831, USA

[2]Department of Physics and Astronomy, Vanderbilt University, Nashville, TN 37235, USA

[3]Interdisciplinary Materials Science Program, Vanderbilt University, Nashville, TN 37235, USA

[4]Department of Physics & Astronomy and California NanoSystems Institute, University of California, Los Angeles, CA 90095, USA

[5]Department of Materials Science and Engineering, North Carolina State University, Raleigh, NC 27695, USA

[6]Department of Materials Science and Engineering, University of California, Berkeley, CA 94720, USA

[7]Department of Electrical and Computer Engineering, Vanderbilt University, Nashville, TN 37235, USA

[8]Department of Mechanical Engineering and Materials Science, Duke University, Durham, NC 27708, USA

*Corresponding authors. Email: kimy6@ornl.gov (Y.-H.K.); harrison.a.walker@vanderbilt.edu (H.A.W.); pantelides@vanderbilt.edu (S.T.P.); chim@ornl.gov (M.C.)

Notice: This manuscript has been authored by UT-Battelle, LLC, under Contract No. DE-AC05-00OR22725 with the U.S. Department of Energy. The United States Government retains and the publisher, by accepting the article for publication, acknowledges that the United States Government retains a non-exclusive, paid-up, irrevocable, world-wide license to publish or re-produce the published form of this manuscript, or allow others to do so, for the United States Government purposes. The Department of Energy will provide public access to these results of federally sponsored research in accordance with the DOE Public Access Plan (http://energy.gov/downloads/doe-public-access-plan).

## Abstract

**Freestanding oxide membranes provide geometric freedom to control interfacial registry and strain through twisting. Although moiré-controlled in-plane polar vortices have been observed in twisted oxide membranes and polar skyrmions predicted in large-angle twisted $SrTiO_3$, three-dimensional polar topologies remain experimentally unexplored. Using depth-resolved atomic-scale electron microscopy and machine-learning atomistic modeling, we reveal spatially registered polar skyrmion lattices in twisted $NaNbO_3$ membranes. The lattices carry opposite topological charges, forming a topologically compensated pair, and exhibit moiré-periodic in-plane polarization winding and depth-dependent out-of-plane polarization. They are associated with a coherent, chemically continuous junction undergoing three-dimensional structural reconstruction: the moiré-locked shear strain reverses across the interface while the lattice rotates through the membrane thickness. These findings establish strongly bonded oxide moirés as a platform for engineering coupled polar textures.**

In complex oxides, coupled lattice, charge, orbital and interfacial interactions can generate non-uniform polar textures, including flux-closure domains (*1*, *2*), vortices (*3*), merons (*4*, *5*) and skyrmions (*6*, *7*). These states provide opportunities to control polarization and topology at the nanoscale, with potential relevance to high-density and multistate electronics (*8*, *9*). Prior studies in oxide films and superlattices have established powerful routes to stabilize and control such textures (*10*, *11*), from polar vortices and skyrmions in $PbTiO_3/SrTiO_3$ superlattices (*3*, *6*) to related textures in $BaTiO_3/SrTiO_3$ superlattices (*12*), $Pb(Zr,Ti)O_3$ (*2*) and $BiFeO_3$ (*13*, *14*) thin films. In these systems, epitaxial coupling to the substrate, which is fixed during growth, is a key parameter governing the formation and structure of the polar state.

Freestanding oxide membranes are released from the substrate-imposed boundary conditions (*15*–*17*) and provide additional geometric freedom for assembling oxide interfaces after growth (*18*). Twisting two membranes while preserving coherent interfacial bonding creates a moiré junction in which the local registry, and hence the interfacial interactions, vary periodically in space (*19*–*21*). Recent studies have revealed moiré-periodic in-plane polar vortices in twisted $BaTiO_3$ (*22*), $SrTiO_3$ (*23*) and $SrRuO_3$ (*24*) membranes, with ferroic order extending beyond the unit-cell length scale. In a related study of twisted $SrTiO_3$, Sandholt et al. (*25*) identified an interfacial screw-dislocation network and experimentally resolved in-plane vortex-antivortex arrays registered to that network rather than to the geometric moiré pattern. Their phase-field calculations predicted an out-of-plane polarization component localized near the interface but did not assign a specific three-dimensional (3D) topology. Other theoretical studies predicted polar skyrmions in twisted $SrTiO_3$ only at large twist angles (36.87° and 22.62°) (*26*, *27*). These experimental and theoretical advances motivate the question of whether specific 3D polar topologies can be realized experimentally in twisted oxide membranes, and whether twist-induced reconstruction extends beyond the interface into the membrane interior, where it may contribute to stabilizing these textures.

Here we study twisted freestanding $NaNbO_3$ membranes (t-NNO). $NaNbO_3$ hosts several closely competing polymorphs, making its structural and polar states sensitive to lattice perturbations (*28*–*30*). In the moiré structure, we find three-dimensional structural reconstruction accompanied by a moiré-locked strain field that reverses across the interface. Our combined analysis further reveals spatially registered three-dimensional polar skyrmion lattices in the two membranes, with opposite topological charges.

## Twist-engineered large moiré strain and coherent interfacial bonding

$NaNbO_3$ films were grown by pulsed laser deposition on $La_{0.7}Sr_{0.3}MnO_3$-buffered $SrTiO_3(001)$ substrates and freestanding membranes were obtained by selectively etching the $La_{0.7}Sr_{0.3}MnO_3$ sacrificial layer (*20*, *21*). Two released membranes were stacked at controlled twist angles ranging from 0.3° to 30.0° and subsequently annealed at 660 °C in $O_2$, yielding t-NNO (fig. S1). Atomic-resolution high-angle annular dark-field (HAADF) imaging resolves the moiré superlattices (Fig. 1, A to D, top), whose periodicity decreases systematically with increasing twist angle. The experimentally measured periodicities ($L$) are approximately 3.4 nm for 6.7°, 1.5 nm for 15.2° and 0.76 nm for 30.0°, in close agreement with the geometric moiré relation $L = a/2\sin(\theta/2)$, where a is the pseudocubic lattice parameter and θ is the twist angle. The corresponding shear strain ($E_{xy}$) maps (Fig. 1, A to D, bottom) exhibit periodic modulations with alternating positive and negative strain cores. These modulations follow the moiré periodicity and are therefore confined to shorter length scales as the twist angle increases.

Given the coupling between polarization and strain gradients through flexoelectricity (*31*, *32*), we quantify the spatial gradients of the $E_{xy}$ fields using directional differentiation (fig. S2), focusing on the 15.2° sample, which exhibits the most clearly resolved and spatially ordered strain texture (Fig. 1C). The $E_{xy}$ field reaches values of up to ~7%, and the average strain-gradient magnitude is ~$2 \times 10^8$ $m^{-1}$ over a characteristic length scale of ~1.5 nm. This gradient is about one order of magnitude larger than the ~$10^7$ $m^{-1}$ gradients reported for twisted $BaTiO_3$ membranes (*22*), which exhibit shear strain of ~3%. An enlarged plan-view HAADF image (Fig. 2A) and the corresponding $E_{xy}$ map (Fig. 2B) of the 15.2° sample reveal alternating positive and negative shear domains arranged in moiré-periodic closed loops around the strain cores. The close correspondence between these features and the moiré superlattice indicates that the $E_{xy}$ field is moiré-locked, that is, spatially pinned to the moiré registry.

To establish whether the two membranes form a coherently bonded junction, we compared the experimental selected-area electron diffraction (SAED) pattern (Fig. 2C) with a simulation based on a rigidly superimposed structural model (Fig. 2D). Although the model reproduces the principal Bragg reflections from both lattices, it does not account for additional experimental reflections, indicating reconstruction beyond rigid geometric superposition. Because SAED integrates through the membrane thickness, we examined the junction directly in cross-section. The interface is uniform over lateral distances exceeding 100 nm (Fig. 2E, top) and remains coherent and atomically flat, without discernible dislocations, defects, or an amorphous interlayer (Fig. 2E, bottom; Fig. 2, F and G). Interfacial screw-dislocation networks (*25*) have been reported in $SrTiO_3$ bilayers twisted by 5° and 10°. In the 15.2° t-NNO junction, we observe a coherent, moiré-locked shear-strain modulation. Uniform Na, Nb and O distributions in energy-dispersive X-ray spectroscopy (EDX) maps, together with complementary electron energy-loss spectroscopy (EELS) measurements, further confirm chemical continuity and exclude a contaminated or chemically defective interlayer (fig. S3). These observations, along with the additional diffraction features that are reproduced by the reconstructed atomistic model described below (fig. S4), indicate that the two membranes are strongly bonded through an atomically coherent and chemically continuous interface, and that the twisted junction is structurally reconstructed rather than a rigid superposition of the two lattices.

**3D structural reconstruction in twisted membranes**

A strongly bonded interface that sustains moiré-locked strain and large strain gradients raises a central question: is the reconstruction confined to the junction, or does it extend through the thickness of both membranes? We address this question using plan-view depth-sectioning HAADF imaging (fig. S5). Figure 3, A to F, shows atomic-resolution images magnified from the depth-sectioning series (left) and the corresponding $E_{xy}$ maps (middle) for focal planes within the top membrane, at the twisted interface and within the bottom membrane of the 15.2° t-NNO sample. These maps show that the $E_{xy}$ field is not confined to the interface but extends through nearly the full thickness of both membranes. Within the top membrane, it is present well away from the interface and increases in magnitude toward it, reaching a maximum at the interface; across the interface, it reverses sign and decays through the bottom membrane to a nearly strain-free state. To facilitate an intuitive understanding of these strain behaviors, pictograms (Fig. 3, A to F, right) depict the corresponding shear distortion of the local unit cell, with '+' ($E_{xy} > 0$; red) and '−' ($E_{xy} < 0$; blue) denoting positive and negative shear textures, respectively.

The depth dependence extends beyond the shear strain itself. Superposing the unit cells extracted along the $E_{xy} = 0$ contours (yellow) reveals a continuous lattice rotation through the thickness (Fig. 3G). Multislice electron ptychography, a computational microscopy technique that reconstructs quantitative phase images from coherent diffraction patterns (*33*, *34*), further confirms that the rotation evolves continuously with depth (movie S1). Tracking the coincidence-site lattice (CSL) position reveals a concomitant lateral shift of the CSL through the thickness (fig. S6). Figure 3H schematically summarizes this depth-dependent structural evolution: the $E_{xy}$ magnitude increases toward the interface, reverses sign across the junction and decreases into the bottom membrane. This evolution is accompanied by progressive lattice rotation and lateral displacement of the strain cores, which extend through nearly the full thickness of the top membrane (~6 nm) and most of the bottom membrane (~4–5 nm, fig. S5 and supplementary text S1). This contrasts with the findings of Sandholt et al. in twisted $SrTiO_3$ (*25*), where structural reconstruction was concentrated within a few atomic planes of the interface, and their phase-field modeling predicted polarization extending approximately 2 nm from the interface. The difference suggests that the depth of moiré-induced structural reconstruction may depend on material-specific lattice instabilities, interfacial coupling, and mechanical boundary conditions.

To identify the origin of this 3D structure, we performed atomistic calculations using the Orb-v3 universal machine-learning interatomic potential (*35*, *36*). We first determined the energetically preferred interfacial configuration by comparing the binding energies of the possible interface terminations, which showed the $NaO/NbO_2$ termination to be the most strongly bound (fig. S7B). Relative to the rigid superposition of the two unrelaxed atomic models (fig. S7A), the relaxed (fig. S7C) and 300 K-averaged (Fig. 3I) structures exhibit a 3D lattice rotation that evolves through the membrane thickness. Furthermore, the strain maps derived from the 300 K-averaged structure at the interfacial planes immediately above and below the twisted interface (hereafter, the upper and lower interfacial planes) reproduce the experimentally observed periodicity, alternating sign and localization of the $E_{xy}$ domains within the moiré cell (Fig. 3, J and K). This agreement indicates that the moiré-locked strain texture originates from reconstruction rather than from the geometric moiré structure alone. Together with the depth-resolved observations (Fig. 3), these results suggest a “twist-induced 3D structural reconstruction” that is not confined to the junction but extends into both membranes, accompanied by depth-dependent shear strain and lattice rotation.

**Coupled 3D polar skyrmion lattices**

The 3D reconstruction provides the structural setting for the polar response. To understand the polar topology, we determined the local polar displacements from the multislice electron ptychographic (MEP) reconstruction of the 15.2° t-NNO sample at planes crossing the twisted interface (Fig. 4, A and B). The resulting maps (Fig. 4, C and D) show moiré-periodic lattices of in-plane polar vortices. The arrows indicate the magnitude and direction of the in-plane polarization, while the color scale denotes the rotational sense of the polarization field (P) as determined from its signed curl ($\omega_z = \partial Py/\partial x - \partial Px/\partial y$). These maps reveal lattices of in-plane vortices whose positions follow the moiré superlattice and hence the moiré-locked $E_{xy}$ field, with the rotational direction of the vortices reversing from counterclockwise (CCW, red) to clockwise (CW, blue) across the interface. For comparison, we examined the polarization and vorticity fields obtained from the 300 K-averaged atomistic model. Their periodicity and sign distribution closely reproduce the experimental maps, including the reversal of the rotational direction at each moiré site across the interface (Fig. 4, E and F). Note that the polarization and vorticity fields were evaluated over the full CSL, folded into symmetry-equivalent moiré subtiles and averaged to improve the signal-to-noise ratio while preserving the moiré symmetry (fig. S8).

While these maps establish the in-plane winding of the polarization textures, this information alone is insufficient to determine their topological class. A vortex and a skyrmion share a similar in-plane winding; they differ in whether the polarization develops an out-of-plane component that reverses between the core and its surroundings, such that the vector field covers the order-parameter sphere (*37*, *38*). Cross-sectional atomic resolution ADF images were analyzed in detail (fig. S9). They show that the polar displacements are not purely in-plane: the polarization direction evolves from upward out-of-plane, through a nearly in-plane orientation, to downward out-of-plane. This indicates that t-NNO contains out-of-plane polarization components of both signs. Together with the moiré-periodic in-plane winding, these observations are consistent with key features of a skyrmion-like texture. However, quantitative determination of the three-dimensional topology from ADF images is hindered by complex vortex-related atomic displacements and projection through a specimen thickness spanning many moiré periods, which obscures the moiré-scale out-of-plane polarization.

To resolve the out-of-plane component within the moiré cell and determine the 3D topology, we derived it from the 300 K-averaged atomistic model, whose in-plane vorticity reproduces the experimentally measured maps. Extracting this component from the model requires a local polarization proxy. At the interface, however, the Nb-O coordination environments are reconstructed, so that the assignment of a unique $NbO_6$ octahedron to each Nb site becomes ambiguous (fig. S10) and the conventional octahedron-based proxy is of limited reliability in this region. We instead use the displacement of each Nb atom relative to the midpoint of the neighboring Na sheets ($u_z$) as a proxy for the out-of-plane polarization (*36*). Combined with the in-plane winding, the resulting maps reveal a pair of polar skyrmion lattices on opposite sides of the interface (Fig. 4, G and H). The polar textures therefore exhibit in-plane winding around a core alongside an out-of-plane component that reverses between the core and its surroundings, so that the vector field wraps the order-parameter sphere and carries a non-zero topological charge (skyrmion number, Q). Because the core polarity and the in-plane rotation reverse together across the interface ($P \rightarrow -P$), the reversal preserves the skyrmion handedness, set by the sign of $P\cdot(\nabla\times P)$ and right-handed in both membranes, whereas Q changes sign.

## Topologically compensated anti-polar skyrmion pairs

To confirm the skyrmion topology, we computed $Q = (1/4\pi) \int n \cdot (\partial n/\partial x \times \partial n/\partial y)\, dx\, dy$, from the unit-vector field $n = P/|P|$ of the moiré-averaged atomistic polar field, integrating over the Voronoi basin of each core (fig. S11). In the top membrane, the two cores within the moiré cell contribute $Q_1 = +1.08$ and $Q_2 = +0.92$, giving a total of $Q = +2$. In the bottom membrane, the corresponding basins yield $Q_1 = -1.05$ and $Q_2 = -0.95$, giving $Q = -2$. The integer totals per moiré cell indicate a lattice of polar skyrmions in each membrane; the deviations of the individual basin charges from unity are consistent with the partition of a continuous periodic field into finite basins. These calculations therefore confirm that the experimentally observed moiré-periodic in-plane winding and out-of-plane polarization components are signatures of polar skyrmion lattices in t-NNO.

Figure 5A visualizes the full 3D polar field obtained by combining the in-plane polarization with the out-of-plane component. Across the twisted interface, the two spatially registered skyrmion lattices reverse their core polarity, rotational direction, and topological charge, whereas the in-plane winding number (+1 for both CW and CCW vortices) and handedness are preserved; the topological charges of the two membranes ($Q = +2$ and $-2$ per moiré cell) therefore cancel, yielding a topologically compensated bilayer. Although each $NaNbO_3$ membrane is ferroelectric (*21*, *29*), this coupled configuration forms a topologically compensated, anti-polar skyrmion pair (Fig. 5B). Here "anti-polar" refers to the opposite core polarity and topological charge of the two lattices, not to the antiferroelectric phase of bulk $NaNbO_3$. This configuration is analogous to antiparallel spin textures in synthetic antiferromagnets (*39–42*), where layers carrying opposite topological charges are coupled through interlayer exchange, although the coupling mechanism and dynamics in t-NNO need not be the same. The resulting state may therefore be regarded as a "synthetic anti-polar skyrmion pair".

The formation of the polar textures in t-NNO is likely a product of the moiré registry and the resulting 3D reconstruction. Such reconstruction may be further facilitated by the rich structural polymorphism of $NaNbO_3$, associated with competing soft phonon modes and oxygen-octahedral rotations and distortions, which could provide a favorable energy landscape for moiré-induced structural rearrangements. This possibility is consistent with the pronounced reconstruction of Nb-O coordination environments found near the twisted interface. In the simulations, Nb-O bonds form across the junction at the subset of interfacial Nb sites where the moiré registry places an O of the opposing layer above an Nb (fig. S10). These bonded sites pin the two lattices, so that the junction is strongly bonded locally even though its binding energy averaged over the CSL is modest (fig. S7B), and the moiré-locked strain field reflects their elastic accommodation. Together with elastic, electrostatic, and polarization-gradient energies, this flexoelectric coupling could contribute to stabilizing the observed skyrmion topology. This route differs from polar textures in $PbTiO_3/SrTiO_3$ (*3*, *6*) superlattices, $BaTiO_3/SrTiO_3$ (*12*) superlattices and $BiFeO_3$ (*13*, *14*) thin films, where the topology is largely prescribed by epitaxial boundary conditions established during growth. The resulting polar textures also differ from the moiré-periodic vortex-antivortex arrays reported in twisted $BaTiO_3$ (*22*), $SrTiO_3$ (*23*) and $SrRuO_3$ (*24*), and the dislocation-associated vortices in twisted $SrTiO_3$ (*25*). Our combined experimental and atomistic modeling results establish polar skyrmion textures in t-NNO, characterized by coupled in-plane polarization winding and out-of-plane reversal, with nonzero, membrane-resolved topological charges. Notably, whereas theoretical studies have predicted polar skyrmions in twisted $SrTiO_3$ only at large twist angles (36.87° and 22.62°) (*26*, *27*), our results demonstrate skyrmion lattices in t-NNO at a substantially smaller twist angle of 15.2°.

In conclusion, combined depth-resolved electron microscopy and atomistic modeling identify a coupled 3D polar-skyrmion state in twisted $NaNbO_3$. The coherently bonded junction hosts spatially registered skyrmion lattices with opposite membrane-resolved charges ($Q = +2$ and $-2$ per moiré cell), while each membrane retains its own non-trivial topology. This paired state is associated with structural reconstruction through both membranes, including a moiré-locked strain field that reverses across the interface and lattice rotation that evolves with depth. These findings establish coherent oxide moiré interfaces as a platform for engineering collective polar topology and motivate exploration of such states for multistate memory, neuromorphic computing and low-power information processing.

## References and Notes

1. Y. L. Tang, Y. L. Zhu, X. L. Ma, A. Y. Borisevich, A. N. Morozovska, E. A. Eliseev, W. Y. Wang, Y. J. Wang, Y. B. Xu, Z. D. Zhang, S. J. Pennycook, Observation of a periodic array of flux-closure quadrants in strained ferroelectric $PbTiO_3$ films. *Science* **348**, 547–551 (2015).
2. C.-L. Jia, K. W. Urban, M. Alexe, D. Hesse, I. Vrejoiu, Direct observation of continuous electric dipole rotation in flux-closure domains in ferroelectric $Pb(Zr,Ti)O_3$. *Science* **331**, 1420–1423 (2011).
3. A. K. Yadav, C. T. Nelson, S. L. Hsu, Z. Hong, J. D. Clarkson, C. M. Schlepütz, A. R. Damodaran, P. Shafer, E. Arenholz, L. R. Dedon, D. Chen, A. Vishwanath, A. M. Minor, L. Q. Chen, J. F. Scott, L. W. Martin, R. Ramesh, Observation of polar vortices in oxide superlattices. *Nature* **530**, 198–201 (2016).
4. Y. J. Wang, Y. P. Feng, Y. L. Zhu, Y. L. Tang, L. X. Yang, M. J. Zou, W. R. Geng, M. J. Han, X. W. Guo, B. Wu, X. L. Ma, Polar meron lattice in strained oxide ferroelectrics. *Nat. Mater.* **19**, 881–886 (2020).
5. Y.-T. Shao, S. Das, Z. Hong, R. Xu, S. Chandrika, F. Gómez-Ortiz, P. García-Fernández, L.-Q. Chen, H. Y. Hwang, J. Junquera, L. W. Martin, R. Ramesh, D. A. Muller, Emergent chirality in a polar meron to skyrmion phase transition. *Nat. Commun.* **14**, 1355 (2023).
6. S. Das, Y. L. Tang, Z. Hong, M. A. P. Gonçalves, M. R. McCarter, C. Klewe, K. X. Nguyen, F. Gómez-Ortiz, P. Shafer, E. Arenholz, V. A. Stoica, S.-L. Hsu, B. Wang, C. Ophus, J. F. Liu, C. T. Nelson, S. Saremi, B. Prasad, A. B. Mei, D. G. Schlom, J. Íñiguez, P. García-Fernández, D. A. Muller, L. Q. Chen, J. Junquera, L. W. Martin, R. Ramesh, Observation of room-temperature polar skyrmions. *Nature* **568**, 368–372 (2019).
7. W. R. Geng, Y. L. Zhu, M. X. Zhu, Y. L. Tang, H. J. Zhao, C. H. Lei, Y. J. Wang, J. H. Wang, R. J. Jiang, S. Z. Liu, X. Y. San, Y. P. Feng, M. J. Zou, X. L. Ma, Dipolar wavevector interference induces a polar skyrmion lattice in strained $BiFeO_3$ films. *Nat. Nanotechnol.* **20**, 366–373 (2025).
8. J. Junquera, Y. Nahas, S. Prokhorenko, L. Bellaiche, J. Íñiguez, D. G. Schlom, L.-Q. Chen, S. Salahuddin, D. A. Muller, L. W. Martin, R. Ramesh, Topological phases in polar oxide nanostructures. *Rev. Mod. Phys.* **95**, 025001 (2023).
9. Z. Hong, Y.-T. Shao, H. Tian, K. Roychowdhury, S. Das, Topological textures in polar materials. *Nat. Phys.* **22**, 1022–1030 (2026).
10. P. Tong, L. Zhou, K. Du, M. Zhang, Y. Sun, T. Sun, Y. Wu, Y. Liu, H. Guo, Z. Hong, Y. Xie, H. Tian, Z. Zhang, Thermal triggering for multi-state switching of polar topologies. *Nat. Phys.* **21**, 464–470 (2025).
11. S. Das, Z. Hong, V. A. Stoica, M. A. P. Gonçalves, Y. T. Shao, E. Parsonnet, E. J. Marksz, S. Saremi, M. R. McCarter, A. Reynoso, C. J. Long, A. M. Hagerstrom, D. Meyers, V. Ravi, B. Prasad, H. Zhou, Z. Zhang, H. Wen, F. Gómez-Ortiz, P. García-Fernández, J. Bokor, J. Íñiguez, J. W. Freeland, N. D. Orloff, J. Junquera, L. Q. Chen, S. Salahuddin, D. A. Muller, L. W. Martin, R. Ramesh, Local negative permittivity and topological phase transition in polar skyrmions. *Nat. Mater.* **20**, 194–201 (2021).

12. S. Estandía, F. Sánchez, M. F. Chisholm, J. Gázquez, Rotational polarization nanotopologies in $BaTiO_3/SrTiO_3$ superlattices. *Nanoscale* **11**, 21275–21283 (2019).
13. N. Balke, B. Winchester, W. Ren, Y. H. Chu, A. N. Morozovska, E. A. Eliseev, M. Huijben, R. K. Vasudevan, P. Maksymovych, J. Britson, S. Jesse, I. Kornev, R. Ramesh, L. Bellaiche, L. Q. Chen, S. V. Kalinin, Enhanced electric conductivity at ferroelectric vortex cores in $BiFeO_3$. *Nat. Phys.* **8**, 81–88 (2012).
14. R. K. Vasudevan, Y.-C. Chen, H.-H. Tai, N. Balke, P. Wu, S. Bhattacharya, L. Q. Chen, Y.-H. Chu, I.-N. Lin, S. V. Kalinin, V. Nagarajan, Exploring topological defects in epitaxial $BiFeO_3$ thin films. *ACS Nano* **5**, 879–887 (2011).
15. Y. Lee, X. Wei, Y. Yu, L. Bhatt, K. Lee, B. H. Goodge, S. P. Harvey, B. Y. Wang, D. A. Muller, L. F. Kourkoutis, W.-S. Lee, S. Raghu, H. Y. Hwang, Synthesis of superconducting freestanding infinite-layer nickelate heterostructures on the millimetre scale. *Nat. Synth.* **4**, 573–581 (2025).
16. S. Choo, S. Varshney, H. Liu, S. Sharma, R. D. James, B. Jalan, From oxide epitaxy to freestanding membranes: Opportunities and challenges. *Sci. Adv.* **10**, eadq8561 (2024).
17. D. Lu, D. J. Baek, S. S. Hong, L. F. Kourkoutis, Y. Hikita, H. Y. Hwang, Synthesis of freestanding single-crystal perovskite films and heterostructures by etching of sacrificial water-soluble layers. *Nat. Mater.* **15**, 1255–1260 (2016).
18. N. Pryds, D.-S. Park, T. S. Jespersen, S. Yun, Twisted oxide membranes: A perspective. *APL Mater.* **12**, 010901 (2024).
19. M.-S. Kim, K. Lee, R. Ishikawa, K. Song, N. A. Shahed, K.-T. Eom, M. S. Rzchowski, E. Y. Tsymbal, N. Shibata, T. Mizoguchi, C.-B. Eom, S.-Y. Choi, Charge disproportionation at twisted $SrTiO_3$ bilayer interface driven by local atomic registry. *ACS Nano* **19**, 39714–39724 (2025).
20. Y.-H. Kim, R. Ghanbari, M.-H. Jung, Y. Zhang, Y.-M. Kim, R. Xu, M. Chi, Establishing atomic coherence in twisted oxide membranes containing volatile elements. *Adv. Mater.* **38**, e74365 (2026).
21. R. Ghanbari, E. Rodrigues, Y.-H. Kim, K. Koons, Y. Li, K. Lebogang, Y. Ding, D. W. Barefoot, Y. Wang, Y. Liu, H. Zhou, M. Chi, R. Xu, Deterministic fabrication of large-area, high-crystallinity oxide moiré superlattices. *ACS Nano* **20**, 20647–20655 (2026).
22. G. Sánchez-Santolino, V. Rouco, S. Puebla, H. Aramberri, V. Zamora, M. Cabero, F. A. Cuellar, C. Munuera, F. Mompean, M. Garcia-Hernandez, A. Castellanos-Gomez, J. Íñiguez, C. Leon, J. Santamaria, A 2D ferroelectric vortex pattern in twisted $BaTiO_3$ freestanding layers. *Nature* **626**, 529–534 (2024).
23. H. Sha, Y. Zhang, Y. Ma, W. Li, W. Yang, J. Cui, Q. Li, H. Huang, R. Yu, Polar vortex hidden in twisted bilayers of paraelectric $SrTiO_3$. *Nat. Commun.* **15**, 10915 (2024).
24. Y. Lun, X. Hu, Q. Ren, U. Saeed, K. Gupta, B. Mundet, I. Pinto-Huguet, J. Santiso, J. Padilla-Pantoja, J. M. Caicedo Roque, M. B. Eriksen, G. Merino, Y. Ma, Q. Li, B. Zhu, G. Tang, D. Pesquera, X. Wang, J. Hong, J. Arbiol, G. Catalan, Polarization vortices in a ferromagnetic metal via twistronics. *Nat. Mater.* 10.1038/s41563-026-02755-8 (2026).
25. W. Sandholt, N. Gauquelin, J. Mangeri, E. Dollekamp, G. Panchal, T. Chennit, A. De Backer, A. Annys, N. Vitaliti, A. R. Insinga, J. M. Hansen, R. Mandal, D. R. Rodrigues, S.

Van Aert, K. I. Wurster, A. Bhowmik, I. E. Castelli, S. B. Simonsen, T. S. Jespersen, R. D. James, B. Jalan, J. Verbeeck, J. M. Garcia Lastra, N. Pryds, Designing dislocation-driven polar vortex networks in twisted perovskites. arXiv:2603.27272 [cond-mat.mtrl-sci] (2026).

26. T. Xu, T. Qian, J. Pang, J. Zhang, S. Li, R. He, J. Wang, T. Shimada, Creating ferroelectricity and ultrahigh-density polar skyrmion in paraelectric perovskite oxide monolayers by moiré engineering. *Research* **8**, 0621 (2025).
27. M. Kono, K. Kasai, Y. Ikeda, T. Xu, S. Minami, T. Shimada, Ultrahigh-density atomic-scale polar skyrmion lattice in $SrTiO_3$ via twist grain boundary engineering. *Adv. Funct. Mater.* **36**, e78177 (2026).
28. R. Machado, M. Sepliarsky, M. G. Stachiotti, Relative phase stability and lattice dynamics of $NaNbO_3$ from first-principles calculations. *Phys. Rev. B* **84**, 134107 (2011).
29. R. Xu, K. J. Crust, V. Harbola, R. Arras, K. Y. Patel, S. Prosandeev, H. Cao, Y.-T. Shao, P. Behera, L. Caretta, W. J. Kim, A. Khandelwal, M. Acharya, M. M. Wang, Y. Liu, E. S. Barnard, A. Raja, L. W. Martin, X. W. Gu, H. Zhou, R. Ramesh, D. A. Muller, L. Bellaiche, H. Y. Hwang, Size-induced ferroelectricity in antiferroelectric oxide membranes. *Adv. Mater.* **35**, 2210562 (2023).
30. R. Ghanbari, H. KP, K. Patel, H. Zhou, T. Zhou, R. Liu, L. Wu, A. Khandelwal, K. J. Crust, S. Hazra, J. Carroll, C. J. G. Meyers, J. Wang, S. Prosandeev, H. Qiao, Y.-H. Kim, Y. Nabei, M. Chi, D. Sun, N. Balke, M. Holt, V. Gopalan, J. E. Spanier, D. A. Muller, L. Bellaiche, H. Y. Hwang, R. Xu, Strain-induced lead-free morphotropic phase boundary. *Nat. Commun.* **16**, 7766 (2025).
31. G. Catalan, A. Lubk, A. H. G. Vlooswijk, E. Snoeck, C. Magen, A. Janssens, G. Rispens, G. Rijnders, D. H. A. Blank, B. Noheda, Flexoelectric rotation of polarization in ferroelectric thin films. *Nat. Mater.* **10**, 963–967 (2011).
32. Q. Li, C. T. Nelson, S.-L. Hsu, A. R. Damodaran, L.-L. Li, A. K. Yadav, M. McCarter, L. W. Martin, R. Ramesh, S. V. Kalinin, Quantification of flexoelectricity in $PbTiO_3/SrTiO_3$ superlattice polar vortices using machine learning and phase-field modeling. *Nat. Commun.* **8**, 1468 (2017).
33. J. Miao, Computational microscopy with coherent diffractive imaging and ptychography. *Nature* **637**, 281–295 (2025).
34. H. Sha, J. Cui, R. Yu, Deep sub-angstrom resolution imaging by electron ptychography with misorientation correction. *Sci. Adv.* **8**, eabn2275 (2022).
35. B. Rhodes, S. Vandenhaute, V. Šimkus, J. Gin, J. Godwin, T. Duignan, M. Neumann, Orb-v3: Atomistic simulation at scale. arXiv:2504.06231 [cond-mat.mtrl-sci] (2025).
36. Materials and methods are available as supplementary materials.
37. N. Nagaosa, Y. Tokura, Topological properties and dynamics of magnetic skyrmions. *Nat. Nanotechnol.* **8**, 899–911 (2013).
38. B. Göbel, I. Mertig, O. A. Tretiakov, Beyond skyrmions: Review and perspectives of alternative magnetic quasiparticles. *Phys. Rep.* **895**, 1–28 (2021).

39. W. Legrand, D. Maccariello, F. Ajejas, S. Collin, A. Vecchiola, K. Bouzehouane, N. Reyren, V. Cros, A. Fert, Room-temperature stabilization of antiferromagnetic skyrmions in synthetic antiferromagnets. *Nat. Mater.* **19**, 34–42 (2020).
40. R. A. Duine, K.-J. Lee, S. S. P. Parkin, M. D. Stiles, Synthetic antiferromagnetic spintronics. *Nat. Phys.* **14**, 217–219 (2018).
41. R. Juge, N. Sisodia, J. Urrestarazu Larrañaga, Q. Zhang, V. T. Pham, K. G. Rana, B. Sarpi, N. Mille, S. Stanescu, R. Belkhou, M.-A. Mawass, N. Novakovic-Marinkovic, F. Kronast, M. Weigand, J. Gräfe, S. Wintz, S. Finizio, J. Raabe, L. Aballe, M. Foerster, M. Belmeguenai, L. D. Buda-Prejbeanu, J. Pelloux-Prayer, J. M. Shaw, H. T. Nembach, L. Ranno, G. Gaudin, O. Boulle, Skyrmions in synthetic antiferromagnets and their nucleation via electrical current and ultra-fast laser illumination. *Nat. Commun.* **13**, 4807 (2022).
42. V. T. Pham, N. Sisodia, I. Di Manici, J. Urrestarazu-Larrañaga, K. Bairagi, J. Pelloux-Prayer, R. Guedas, L. D. Buda-Prejbeanu, S. Auffret, A. Locatelli, T. O. Menteş, S. Pizzini, P. Kumar, A. Finco, V. Jacques, G. Gaudin, O. Boulle, Fast current-induced skyrmion motion in synthetic antiferromagnets. *Science* **384**, 307–312 (2024).

**Acknowledgments**

**Funding:** This experimental work was supported by the U.S. Department of Energy (DOE), Office of Science, Basic Energy Sciences (BES), Materials Sciences and Engineering Division (analysis of structure, chemistry, strain, and polarization behaviors, electron microscopy). Characterization research was conducted as part of a user project at the Center for Nanophase Materials Sciences (CNMS), which is a US Department of Energy, Office of Science User Facility at Oak Ridge National Laboratory. H.A.W. and S.T.P. were supported by the U.S. Department of Energy, Office of Basic Energy Sciences (DOE-BES), Division of Materials Sciences and Engineering grant DE-FG02-09ER46554, the U.S. Department of Energy, Office of Science User Facility, and by the McMinn Endowment. H.A.W. and S.T.P. acknowledge computing resources from the National Energy Research Scientific Computing Center (NERSC), a U.S. Department of Energy Office of Science User Facility located at Lawrence Berkeley National Laboratory, operated under Contract No. DE-AC02-05CH11231. This material is based upon work supported by the National Science Foundation Graduate Research Fellowship Program under Grant No. 2444112 (H.A.W.). Any opinions, findings, and conclusions or recommendations expressed in this material are those of the author(s) and do not necessarily reflect the views of the National Science Foundation. R.G. and R.X. acknowledge support from the Army Research Office under award No. W911NF-25-1-0201, the National Science Foundation (NSF) under award No. DMR-2442399, the American Chemical Society Petroleum Research Fund under award No. 68244-DNI10, and the support from Scialog grant #SA-QMI-2025-097c from Research Corporation for Science Advancement. J.M. acknowledges support from the U.S. Air Force Office of Scientific Research MURI program under Award No. FA9550-25-1-0334.

**Author contributions:** Y.-H.K. and M.C. conceived the work. Y.-H.K. conducted all microscopy experiments and data analysis. H.S. and J.M. performed the multislice electron ptychography reconstruction. H.A.W. and S.T.P. conducted the atomistic calculations and related analyses. R.G., E.R. and R.X. prepared the twisted freestanding membranes. H.N. contributed to TEM specimen preparation. Y.-H.K., R.R., R.X., H.A.W., S.T.P. and M.C. contributed to interpretation of the results. Y.-H.K. and H.A.W. wrote the initial manuscript, with contributions from R.X., S.T.P. and M.C. All authors participated in editing the manuscript.

**Competing interests:** The authors declare that they have no competing interests.

**Data and materials availability:** All data are available in the main text or the supplementary materials. The source codes used in this study are available from the corresponding authors upon reasonable request and for non-commercial use.

**Supplementary Materials**

Materials and Methods
Figs. S1 to S12
Supplementary Text
References
Movie S1

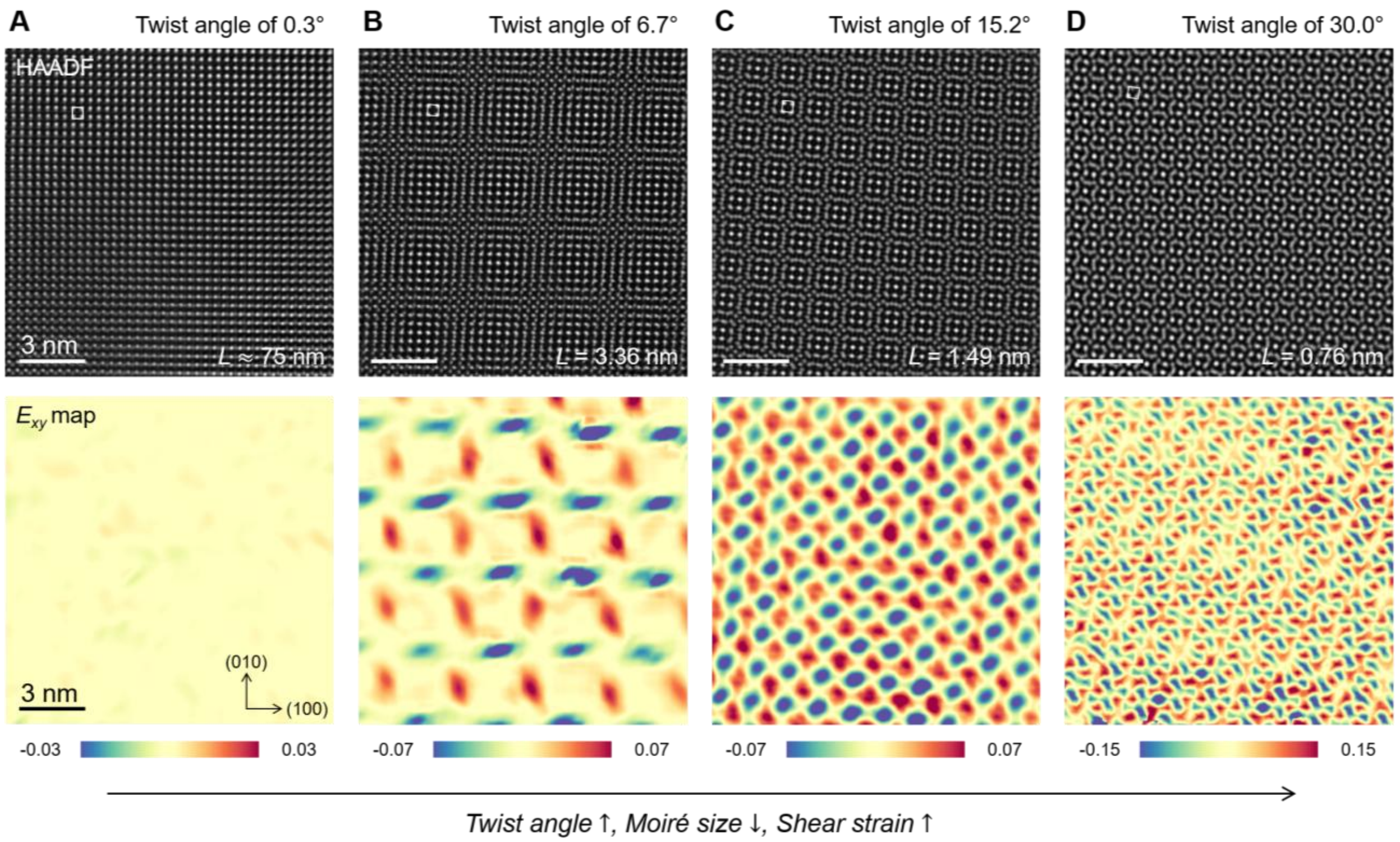

**Fig. 1. Twist-angle-controlled shear strain engineering in twisted $NaNbO_3$ membranes (t-NNO).** (**A** to **D**) Atomic-resolution high-angle annular dark-field (HAADF) images and corresponding shear strain ($E_{xy}$) maps, the shear component referenced to the undistorted lattice (white box), as a function of twist angle: θ = 0.3° (A), 6.7° (B), 15.2° (C), and 30.0° (D). With increasing twist angle, the moiré periodicity decreases while the $E_{xy}$ becomes more localized, concentrating strain variations over shorter length scales and amplifying the strain magnitude. Scale bars are 3 nm. For θ = 0.3° (A), the moiré period expected from the geometric relation exceeds the field of view; the value ($L \approx 75$ nm) shown is calculated rather than measured.

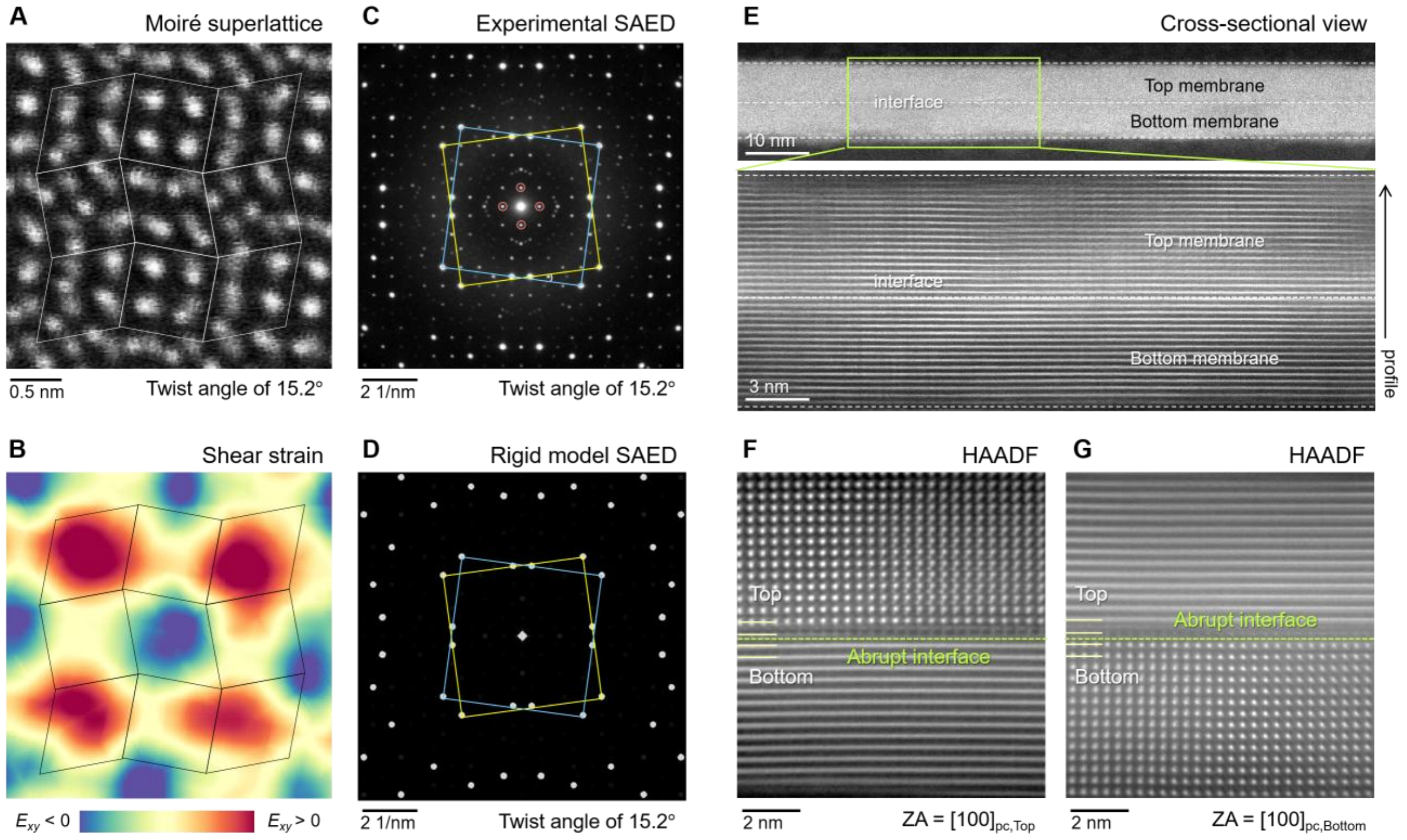

**Fig. 2. Moiré-locked strain field with a coherent, reconstructed interface in twisted $NaNbO_3$ membranes (t-NNO).** (**A**) Enlarged plan-view atomic-resolution HAADF image of the 15.2° t-NNO. (**B**) Corresponding strain ($E_{xy}$) map, showing spatially periodic, alternating positive (red; $E_{xy} > 0$) and negative (blue; $E_{xy} < 0$) strain domains. The oppositely signed strain regions form closed loops around the strain cores. (**C**) Selected-area electron diffraction (SAED) pattern of the 15.2° t-NNO. (**D**) Simulated SAED pattern calculated from a rigidly superimposed structural model with the same nominal twist angle. Differences between the patterns indicate that the rigid model does not fully describe the observed structure and are consistent with structural reconstruction in the twisted heterostructure. (**E**) Cross-sectional ADF images of the twisted junction. The low-magnification image (top) shows a laterally uniform interface extending over more than 100 nm, whereas the lattice-fringe image (bottom) reveals an abrupt and coherent junction without discernible misfit dislocations or amorphous gaps. (**F** and **G**) Representative cross-sectional atomic-resolution HAADF images recorded with the top (F) or bottom (G) membrane on zone axis, showing the atomically flat twisted interface.

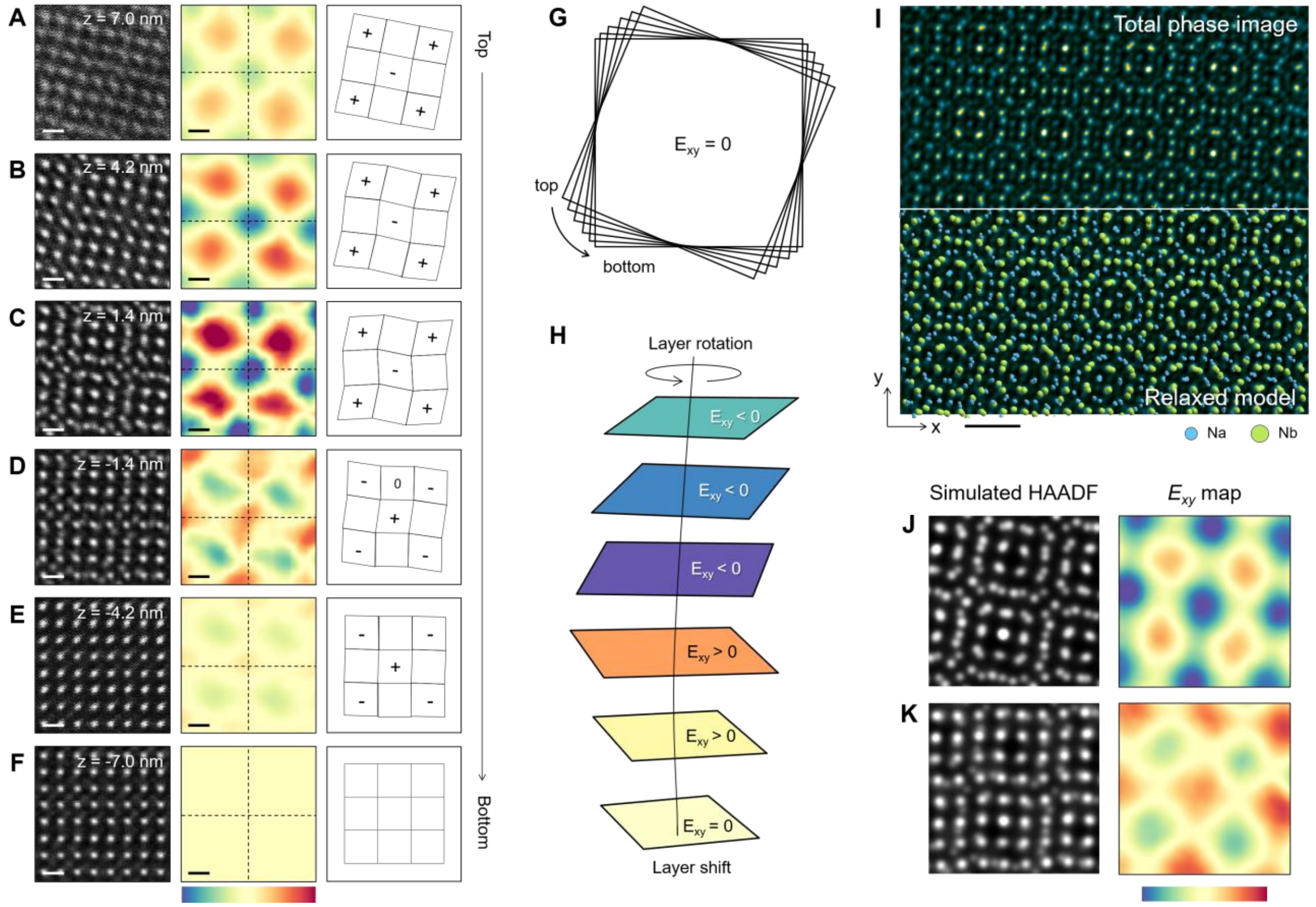

**Fig. 3. Three-dimensional (3D) shear strain field in twisted $NaNbO_3$ membranes (t-NNO).** (**A** to **F**) Plan-view HAADF images (left) and corresponding shear strain ($E_{xy}$) maps (middle), with pictograms of the local shear distortion (right), acquired at nominal focal positions corresponding to the top membrane (A and B), the twisted interface (C and D), and the bottom membrane (E and F). Scale bars are 0.5 nm and the color scale ranges from −0.07 to 0.07. The depth-resolved maps reveal a systematic evolution of the strain field across the membrane thickness: proceeding from the surface toward the twisted interface, the $E_{xy}$ amplitude intensifies, reaching its maximum magnitude and spatial periodicity at the interface. Upon traversing the interface, the $E_{xy}$ undergoes a sign reversal and decays into a nearly strain-free state in the bottom membrane, demonstrating that the twist-induced elastic field is inherently 3D. (**G**) Schematic overlay of unit cells extracted along $E_{xy} = 0$ contours, illustrating a sequential lattice rotation of the unit cell orientation from top to bottom membrane. (**H**) 3D schematic summarizing the depth-dependent evolution of $E_{xy}$, incorporating the layer rotation with a subtle offset of the strain cores, the variation in strain magnitude, and the reversal of the rotational sense of the strain vortices across the membrane thickness. The sign given for each layer denotes $E_{xy}$ at the center of the moiré cell (intersection of the dashed lines in A to F). (**I**) Total phase image of ptychographic reconstruction with simulated structural model averaged from 300 K molecular dynamics, showing agreement between the predicted and observed structure. Scale bar is 1 nm. (**J** and **K**) Simulated HAADF images (left) of the 300 K-averaged structural model at the planes immediately above (J) and below (K) the twisted interface, and the corresponding $E_{xy}$ maps (right).

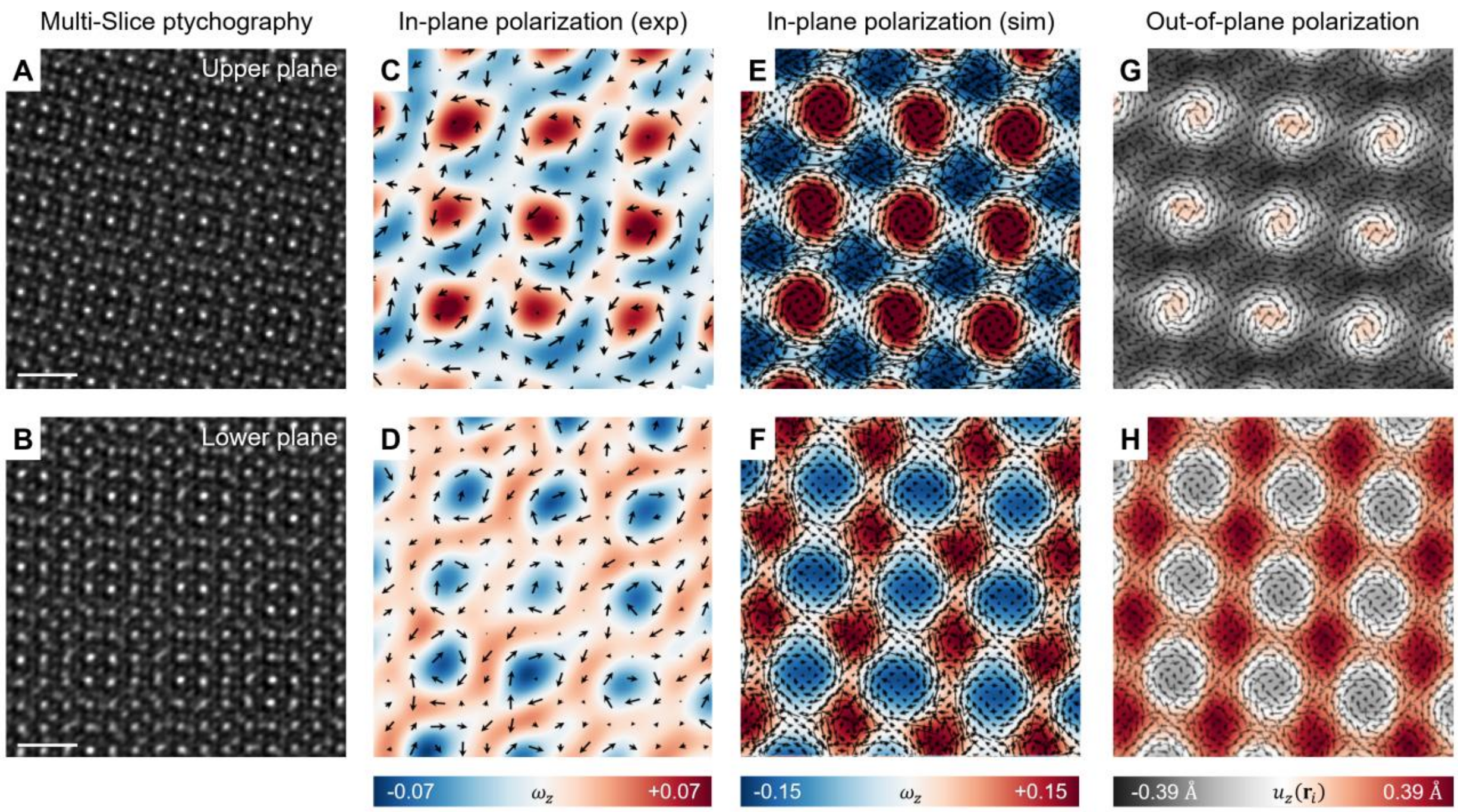

**Fig. 4. Polar skyrmion lattices of opposite topological charge coupled across the twisted $NaNbO_3$ interface (t-NNO).** (**A** and **B**) Ptychographic reconstructions acquired at upper interfacial plane in the top membrane (A) and lower interfacial plane in the bottom membrane (B), respectively. Scale bars are 1 nm. (**C** and **D**) In-plane polarization and vorticity maps obtained at the focal planes. Arrows indicate the magnitude and direction of the in-plane polarization, while the color scale represents vorticity, with clockwise (CW, blue) and counterclockwise (CCW, red) rotation. (**E** and **F**) In-plane polarization and vorticity maps calculated from the 300 K-averaged atomistic model at the upper (E) and lower (F) interfacial planes, respectively. (**G** and **H**) Out-of-plane polarization from the 300 K-averaged atomistic model associated with the skyrmion textures at the upper (G) and lower (H) interfacial planes, respectively. The reversal of the rotation across the interface, together with the variation of the out-of-plane polarization across the in-plane vortex texture, is characteristic of coupled polar skyrmions.

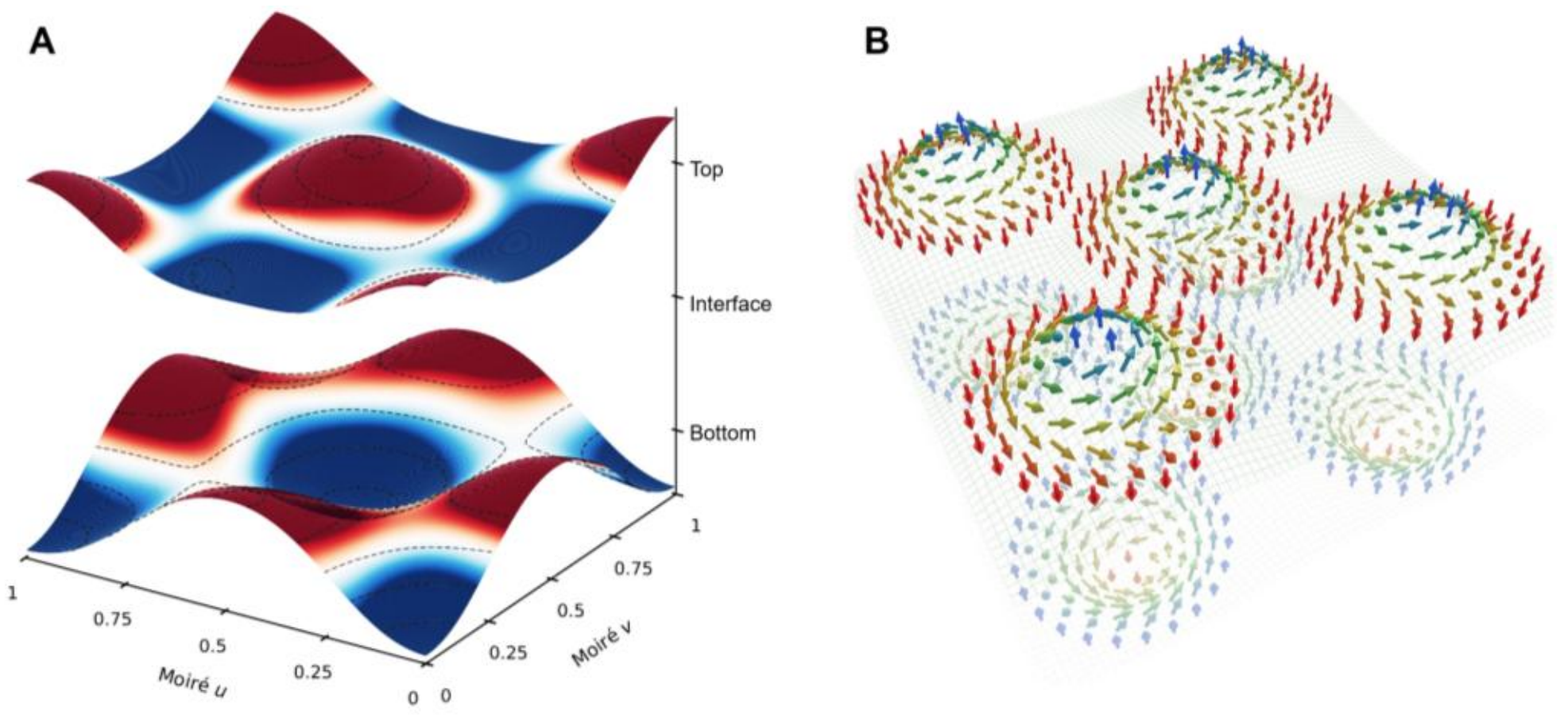

**Fig. 5. Polar skyrmion textures across the twisted $NaNbO_3$ interface.** (**A**) 3D representation of the interfacial polar topology, revealing a lattice of polar skyrmions in each membrane. The color encodes the in-plane vorticity, whereas the surface height represents the out-of-plane polarization, magnified by a factor of five for clarity. Across the reconstructed interface, both the in-plane vorticity and the out-of-plane polarization reverse sign ($P \to -P$). This reverses the skyrmion number ($Q = +2 \to -2$ per moiré cell) while preserving the handedness (sign of $P\cdot(\nabla\times P)$) and the winding number, giving homochiral skyrmion lattices of opposite topological charge on the two sides of the twisted junction. (**B**) Schematic of synthetic anti-polar skyrmion pairs.